\documentclass[twocolumn,aps,prl,reprint,groupedaddress,showpacs,nofootinbib]{revtex4-1}
\usepackage{dcolumn}
\usepackage{bm}
\usepackage{amsmath}

\input{epsf}

\begin{document}
\title{Spin structure and spin magnetic susceptibility of two-dimensional Wigner clusters}

\author{Mehrdad M. Mahmoodian$^{1,2}$}
\email{m.makhmudian1@g.nsu.ru}

\author{M.M. Mahmoodian$^{1,2}$}
\email{mahmood@isp.nsc.ru}

\author{M.V. Entin$^{1}$}
\email{entin@isp.nsc.ru}

\affiliation{$^{1}$ Rzhanov Institute of Semiconductor Physics, Siberian Branch, Russian Academy of Sciences, Novosibirsk, 630090 Russia}
\affiliation{$^{2}$ Novosibirsk State University, Novosibirsk, 630090 Russia}

\date{\today}

\begin{abstract}
Spin states of two-dimensional Wigner clusters are considered at low temperatures, when all electrons are in ground coordinate states. The spin subsystem behavior is determined by antiferromagnetic exchange integrals. The spin states in such a system in the presence of a magnetic field are described in terms of the Ising model. The spin structure, correlation function and magnetic susceptibility of the cluster are found by computer simulations. It is shown that the spin susceptibility experiences oscillations with respect to the magnetic field, owing to the magnetoinduced spin subsystem rearrangements.
\end{abstract}


\maketitle
\section{Introduction}
Two-dimensional Wigner lattices have been the subject of study since the 1970's ~\cite{chap, platzman, grimes}. In these systems the Coulomb repulsion is compensated by opposite charges separated by an insulating layer from a 2D quantum well. The classical electrons occupy the potential energy minima, and quantum electrons form wave functions near the minima. If the quantization energy becomes large, the electron lattice melts and the system converts to electron gas.

At a low temperature and in the absence of quantization, the electrons are arranged in such a way as to minimize their potential energy. Different aspects of the 2D Wigner lattice and 2D Wigner cluster (2DWC) have been studied, for example, the energy of clusters, plasmon spectrum, melting phase transition, and the magnetic field action.

The theoretical approaches to the Wigner lattices study are separated by the classical or quantum consideration neglecting an accounting for spins. Besides, the exchange interaction is weaker than the Coulomb one. However, in the case of small differences between the ordering minima, the exchange interaction may affect the coordinate ordering.

The fine tuning of 2DWC states by the confining potential and magnetic field gives rise to their possible applications in electronics \cite{grabert}, lasers \cite{arakawa,kirst} and quantum computing \cite{loss}.

A specific case of Wigner lattices is the systems with few electrons. In these 2DWCs electrons are kept from running away by an external potential both near the surface and along it.

The  two-stage 2DWC melting is investigated in \cite{bed-peet, lozo1}, and the electron vibration spectrum is considered in \cite{schwei}.

There is a class of systems which should be mentioned, namely, 2D lattices of ions trapped under the superfluid He surface (see {\it e.g.} \cite{vinen}). In this systems the effective mass is much larger than the electron mass, and that contributes to ion condensation.

The experimental realization of the 2DWC in semiconductors met serious difficulties due to a low-needed electron density when the disorder usually overweights the e-e interaction. However, in the recent papers, such systems have been realized as a 1D Wigner cluster with a string-zigzag transition \cite{shapir, ho}, the Wigner crystal in a monolayer semiconductor \cite{smol} and bilayer Wigner crystals in transition metal dichalcogenide heterostructures \cite{zhou}.

Usually, electrons are considered as structureless particles. The coordinate ordering of electrons in a cluster does not directly affect their spin degree of freedom. However, the spin structure of 2DWC is important for its thermodynamic properties.

There are a lot of papers, including reviews, devoted to the study of 2DWC \cite{yan,li,tav-peet,rontani,kou,mak,ghosal,li1,reimann,reimann1,bed-peet,bolt}. The problem of exact Hamiltonian diagonalization is very time-consuming. Just for this reason, the problem was solved for small $n\sim 4$ electron numbers (see {\it e.g.} for the 2D two-center oscillator \cite{li} and 2D harmonic confining potential \cite{tav-peet} with $n=4$, and for $n\leq 8$ \cite{rontani}).

The Ising model is a most popular formulation for the phase transition theory. The 2DWC with a large $n$ contains many aspects of the 2D Ising model, but differs in the finite size and lattice irregularity. Besides, being applied to the 2DWC, the spin ordering of any $n$ is specific.

There are extended studies of Coulomb blockade in few-electron quantum dots (see, {\it e.g.}, review \cite{kou}). The approach of the Coulomb blockade differs from the 2DWC in the smearing of the electron density inside the dot. In this approximation, unlike 2DWC, the correlation between electrons is neglected or assumed to be weak and perturbative.

The review \cite{mak} deals with few interacting quantum electrons in the circular-symmetric parabolic quantum well. The spin ordering is also included. The quantum approach strongly complicates the consideration. As a result, the consideration is limited by $n<7$. This approach does not permit to include the shell and spatial-periodic structure of the cluster.

The parabolic quantum dot with up to 20 electrons was studied using trial wave functions \cite{ghosal}. The confinement strongly affects the correlations due to the broken translational symmetry which results in the electron lattice localization. The exact quantum problem for 3 electrons in an asymmetric parabolic well is considered in \cite{li1}. The calculation limiting yields no ways to extend the electron number essentially.

The spin structure of quasiclassical 2DWC was studied in \cite{reimann,reimann1} for a 1D electron chain.

The results on the electron states in 2D quantum dots and rings are reviewed in \cite{reimann}. Within the reviewed papers concerning the classical ordering, there are \cite{bed-peet,bolt}. The classical configurations with geometric shells ($n_1$,$n_2$,\ldots) of electrons in 2DWC with different electron numbers were obtained in \cite{bed-peet,bolt,weJPCS}. These papers have differences for some electron numbers. In particular, for $n=6$ there are configurations (1,5) \cite{bolt} or (3,3) \cite{bed-peet}, for $n=10$, configuration (3,7) \cite{bed-peet,weJETPL}, or (2,8) \cite{bolt}. These differences originate from very small differences in energy minima corresponding to 2DWC ''isomers''. The isomery leads to an easy melting of the structure at a low temperature.

\begin{figure}
\centerline{\epsfysize=1.3cm\epsfbox{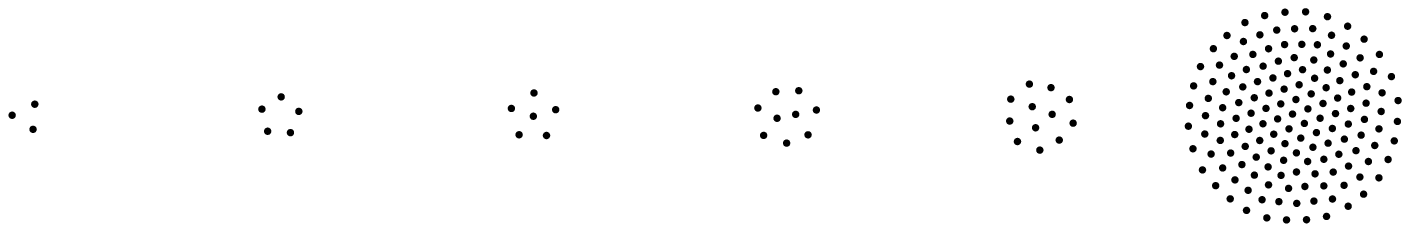}}
\vspace{0.25cm}
\centerline{\epsfysize=0.14cm\epsfbox{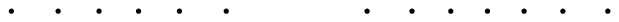}}
\caption{Characteristic structures of the lowest-energy 2DWC for a circular symmetric well ($b=1$) with 3,5,6,9,12,150 electrons and strongly asymmetric ($b=20$) clusters with 6 and 7 electrons. The electron density drops near the cluster edges.}\label{fig1}
\end{figure}

We studied the structure of classical 2DWC in an axially symmetric and asymmetric parabolic potential $k(x^2+b^2 y^2)$ \cite{weJPCS}. Inside the cluster, the electrons form a distorted triangular lattice (see Fig.~\ref{fig1}). The competition between the border and internal energies leads to the polycrystallicity of the inner part of the cluster and to the ordered arrangement of boundary electrons, together with the presence of topological defects inside the cluster. The structure of the classical 2DWC has weak difference from the structure obtained by quantum calculations (for example, in \cite{mak}). The 2DWC rotation under the action of alternating magnetic field was considered in \cite{weJETPL}.

Here we will concentrate on multi-electron clusters. Even if the exchange interaction proofs weak, as compared with the spin-free Hamiltonian, the Heisenberg exchange Hamiltonian is a $2^n\times 2^n$ matrix which dimensionality exponentially grows with $n$. In this case, a simpler Ising Hamiltonian essentially simplifies the problem of spin ordering, conserving the main formulation features. Our purpose is to study the spin structure and the magnetic moment, and susceptibility of electrons in symmetric and asymmetric parabolic quantum wells. We shall consider the many-electron 2DWC based on the classical ordering and then include the exchange interaction to study the spin ordering based on the Ising model. Such consideration of 2DWC is justified by a higher simplicity that permits to expand the considered electron numbers.

Note that the Ising lattices have been intensively studied before. However, many factors differ the 2DWC from regular lattices. In the 2DWC the lattice is frustrated, finite, contains polycrystalline blocks and topological defects. It has also a different density inside and at the cluster boundary. All these factors affect the magnetic behavior.

The article is structured as follows. First, we describe the spatial structure of 2DWCs. Then the exchange interaction of electron spins at the lower coordinate energy level will be described. After that, the Ising model, which determines the statistics of spins, spin structure and susceptibility, will be formulated. Next, we will describe the results of modeling the spin structure of one-dimensional and circular clusters, and the dependence of the spin magnetic susceptibility on the magnetic field. After that, we shall calculate the spin correlation functions at a low temperature and zero magnetic field. Finally, we will discuss the results obtained. The Appendix contains the discussion of the heavy-hole system in a 2D semiconductor quantum well where the Ising model is adequate.

\section{Clusters under study}
We consider free electrons near a flat surface placed in a potential well $k(x^2+b^2y^2)/2$, ($k>0$) and experiencing Coulomb repulsion $\sum\limits_{i>j}\frac{e^2}{\epsilon|{\bf r}_i-{\bf r}_j|}$, where $e$ is the electron charge, $\epsilon$ is the dielectric constant of the surrounding medium, $b$ is the parameter of potential well anisotropy. Further we shall use the dimensionless coordinates by scaling ${\bf r}_i\to L{\bf r}_i$, $L=(2e^2/\epsilon k)^{1/3}$. The energy, measured in unites $E_0=(ke^4/2\epsilon^2)^{1/3}$, reads
\begin{eqnarray}\label{f0}
H=\sum\limits_{i>j}\frac{1}{|{\bf r}_i-{\bf r}_j|}+\sum\limits_i\left(x_i^2+b^2y_i^2\right).
\end{eqnarray}
The asymmetry of the wells determines the corresponding asymmetry of the two-dimensional cluster. For $b=1$, the clusters, on the average, have a circular symmetry; for $b\gg 1 $ or $b\ll 1$, the cluster becomes elongated, and, in the limit - one-dimensional. At low temperatures, electrons populate the energy minima composing a quasi-periodic lattice that minimizes the cluster energy.

The estimate of a 2DWC size with $n$ electrons yields $R\sim(ne^2/\epsilon k)^{1/3}$, distance between electrons $R/\sqrt{n}$ and characteristic electron density $n/R^2$~\cite{weJETPL}.

To find the spin structure, we should take into account the electron quantization. The wave functions of one electron, while the others are immobile, are localized near the bottoms of the wells and have energies $\varepsilon_0=\hbar(k/m_e)^{1/2}n^{1/4}$, where $m_e$ is the electron mass.

The spin ordering is determined by the spin Hamiltonian which will be studied based on the electron coordinates found from the Hamiltonian Eq.~(\ref{f0}). We neglect the effect of spin interaction on the cluster structure. Besides, we assume that all electrons are in the ground coordinate state.

The ground-state wave function of the $i$-th electron decreases as $\psi({\bf r}-{\bf r}_i)\propto\exp(-\alpha|{\bf r}_i-{\bf r}|)$, where $\alpha\sim \sqrt{m_eE_0}Ln^{1/12}/\hbar$. The exchange interaction between the spins of two electrons in the ground state is determined by the overlap integral between the coordinate wave functions $J_{ij}=J_0\exp(-\alpha|{\bf r}_i-{\bf r}_j|)$ and has an antiferromagnetic character. Here and below we neglect the difference between the states of different electrons.

Note that we used the dependence model of exchange integrals $\propto \exp(-\alpha r_{ij})$. This exponential behavior is characteristic for a single-electron wave function at a long distance from the minimum. In fact, this dependence is not universal and does not take into account either the lattice, or the magnetic field. In the presence of the magnetic field the long-range dependence is replaced by $\propto \exp(-r_{ij}^2 /a_B^2)$, where $a_B=\sqrt{c\hbar/eB}$ is a magnetic length, $B$ is magnetic field, $c$ is the light velocity. This dependence leads to the magnetic field-controlled exchange interaction. But, for $ r_{ij}/a_B^2 \ll \alpha$, the magnetic field is inessential for $J_{ij}$.

We shall use the Ising model for the description of spin ordering. This is a commonly used model in the phase transitions theory. The Ising model applicability can be approved for the hole systems in usual semiconductors (see the Appendix). In a magnetic field $B$ the spin energy $E$ reads
\begin{eqnarray}\label{f1}
E=\sum J_{ij}\sigma_i\sigma_j-h\sum\sigma_i,~~~\sigma_i=\pm1,
\end{eqnarray}
where, $h=2\mu_BB/E_0$ and $\mu_B$ is Bohr magneton. The statistical sum is
\begin{eqnarray}\label{f3}
Z=\sum_{\{\sigma_i=\pm 1\}}\exp\left(-\beta E\right).
\end{eqnarray}
Here $\beta=E_0/k_BT$, $T$ is the temperature and $k_B$ is the Boltzmann constant. The quantity $Z$ determines all thermodynamic properties of the system, in particular, the induced magnetic moment $M$ and the spin magnetic susceptibility $\chi$:
\begin{eqnarray}\label{f2}
M=\frac{1}{\beta}\frac{\partial\ln Z}{\partial h},~~~\chi=\frac{\partial M}{\partial h}.
\end{eqnarray}

We performed the numerical calculations based on Eqs.~(\ref{f1}-\ref{f2}).

\section{Magnetic moment}
\begin{figure}[ht]
\centerline{\epsfysize=5.5cm\epsfbox{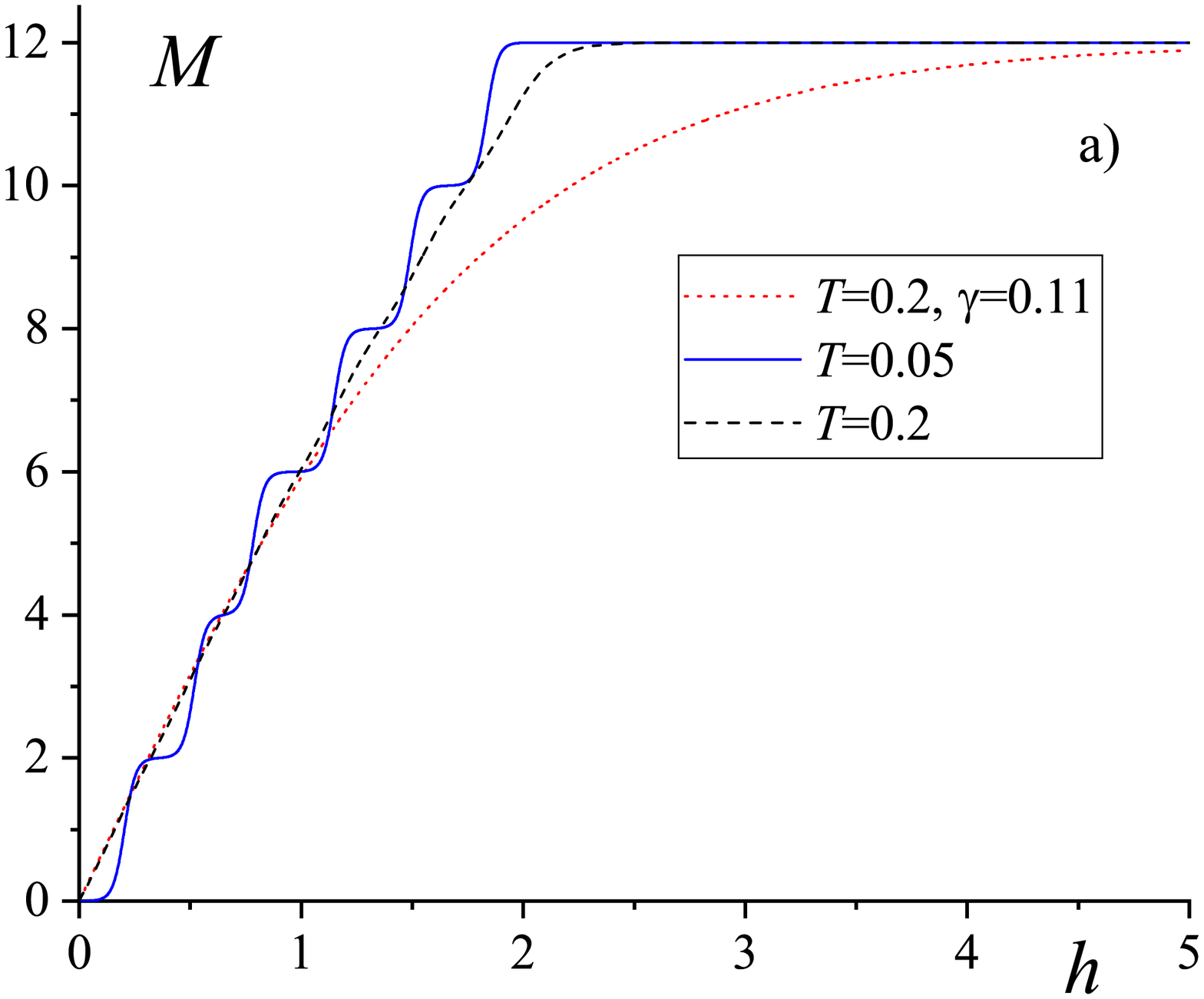}}
\centerline{\epsfysize=5.5cm\epsfbox{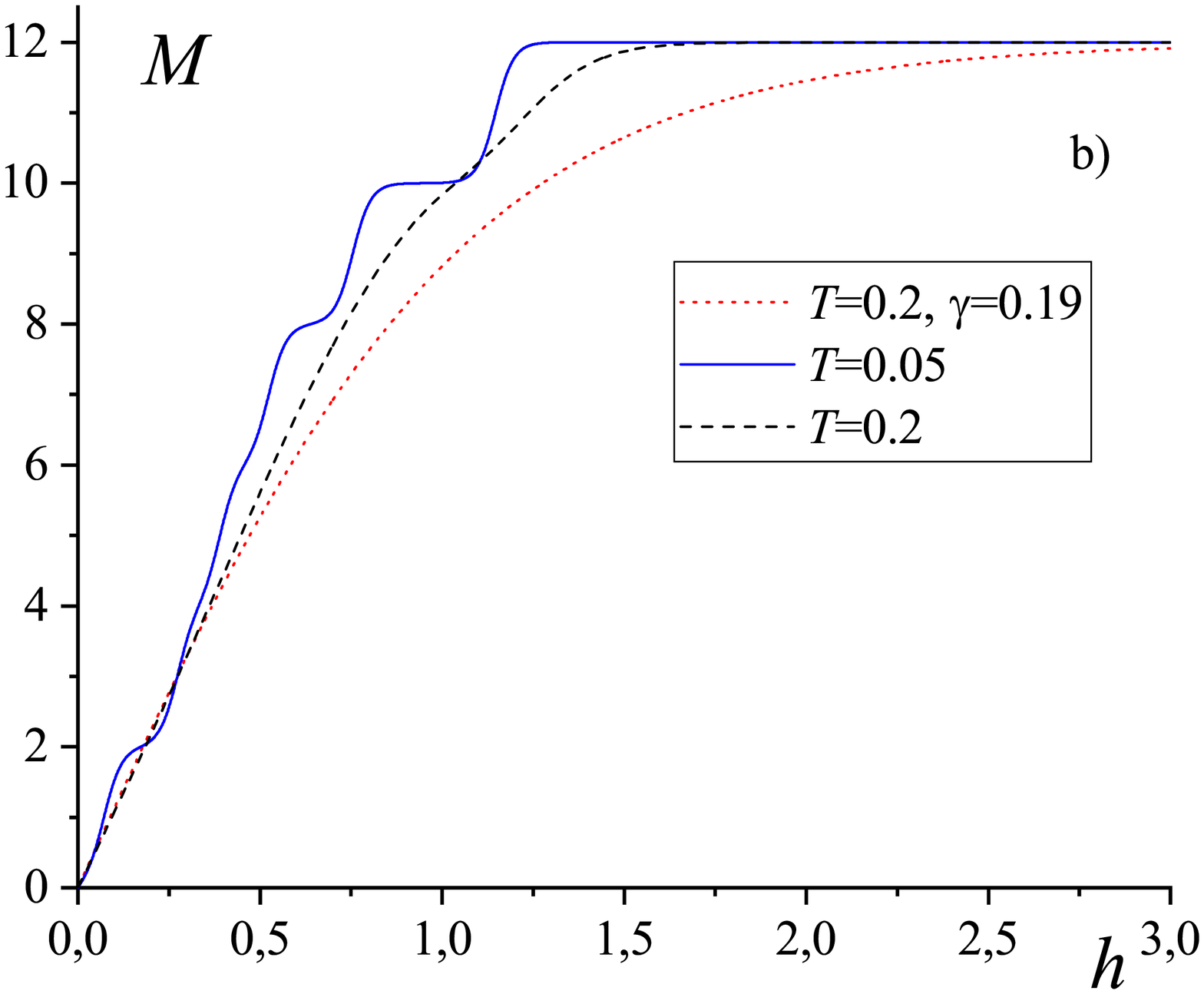}}
\caption{The magnetic spin momentum of linear (a) and round (b) clusters with $n=12$ electrons {\it versus} magnetic field at the relatively high temperature $T=0.2$ (black, dashed) and low temperature $T=0.05$ (blue, solid).}\label{fig2}
\end{figure}

The simulations of the magnetic moment {\it versus} magnetic field for linear and round clusters is presented in Fig.~\ref{fig2} (blue solid and black dashed lines). The magnetic momentum of 2DWC grows with the magnetic field up to the maximal value. The simulations show that the magnetic momentum dependence on $h$ is essentially different from the Curie-Weiss model
$$M(h)=M_0\tanh(\gamma\beta h).$$
At high temperatures (black dashed lines), the growth is more linear than it can be expected from the Curie-Weiss model (red dotted lines). We see that $M(h)\propto~h$ at small~$h$. Then this dependence abruptly goes to $M(h)=M_0$. Another important property of the $M(h)$ dependence is the presence of multiple steps. The steps are explained by the sequential dispairing of spin pairs by the magnetic field. This result is consistent with the periodic peak structure of $\chi(h)$.

\section{Spin susceptibility}
\begin{figure}[ht]
\centerline{\epsfysize=5.5cm\epsfbox{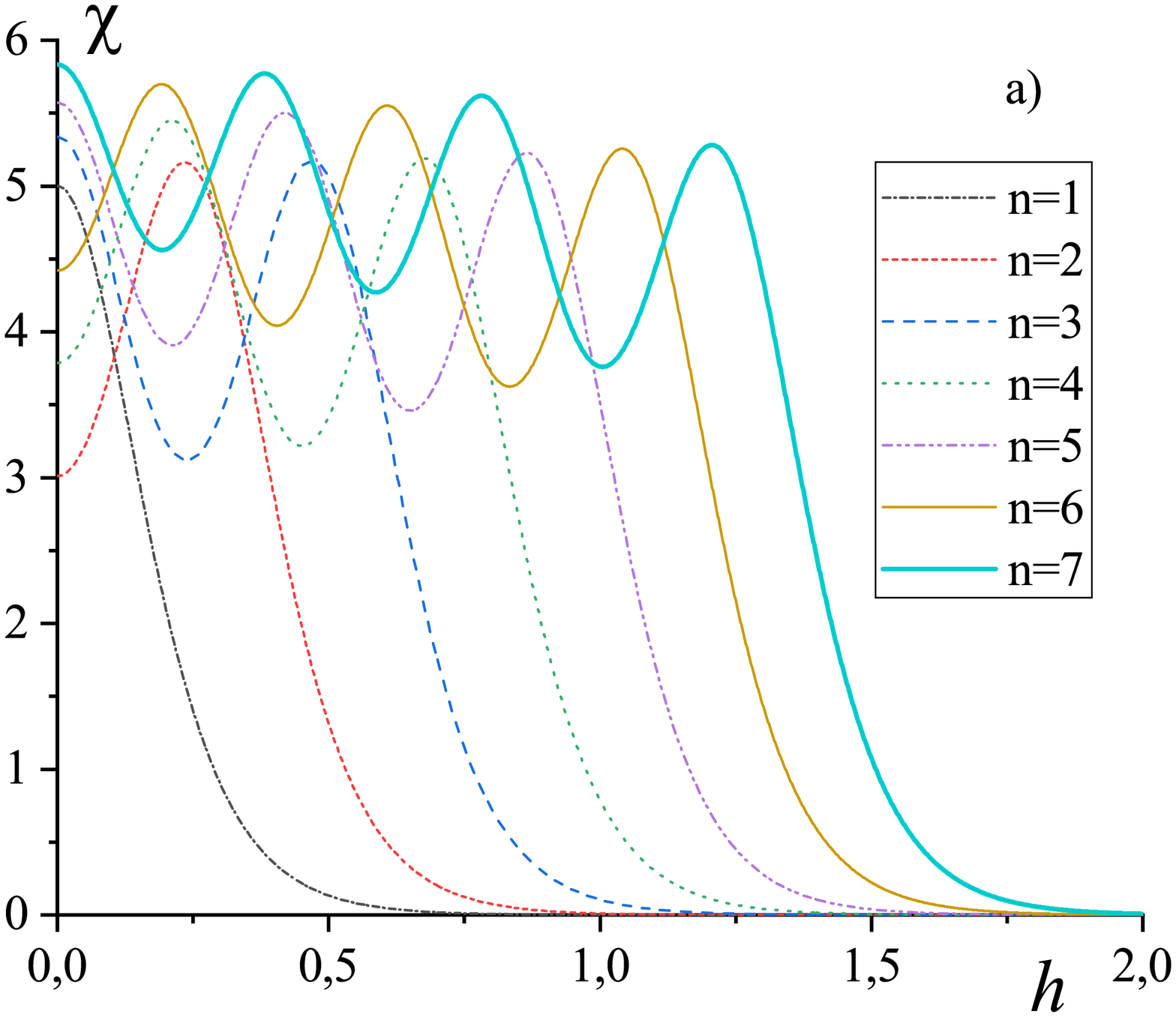}}
\centerline{\epsfysize=5.5cm\epsfbox{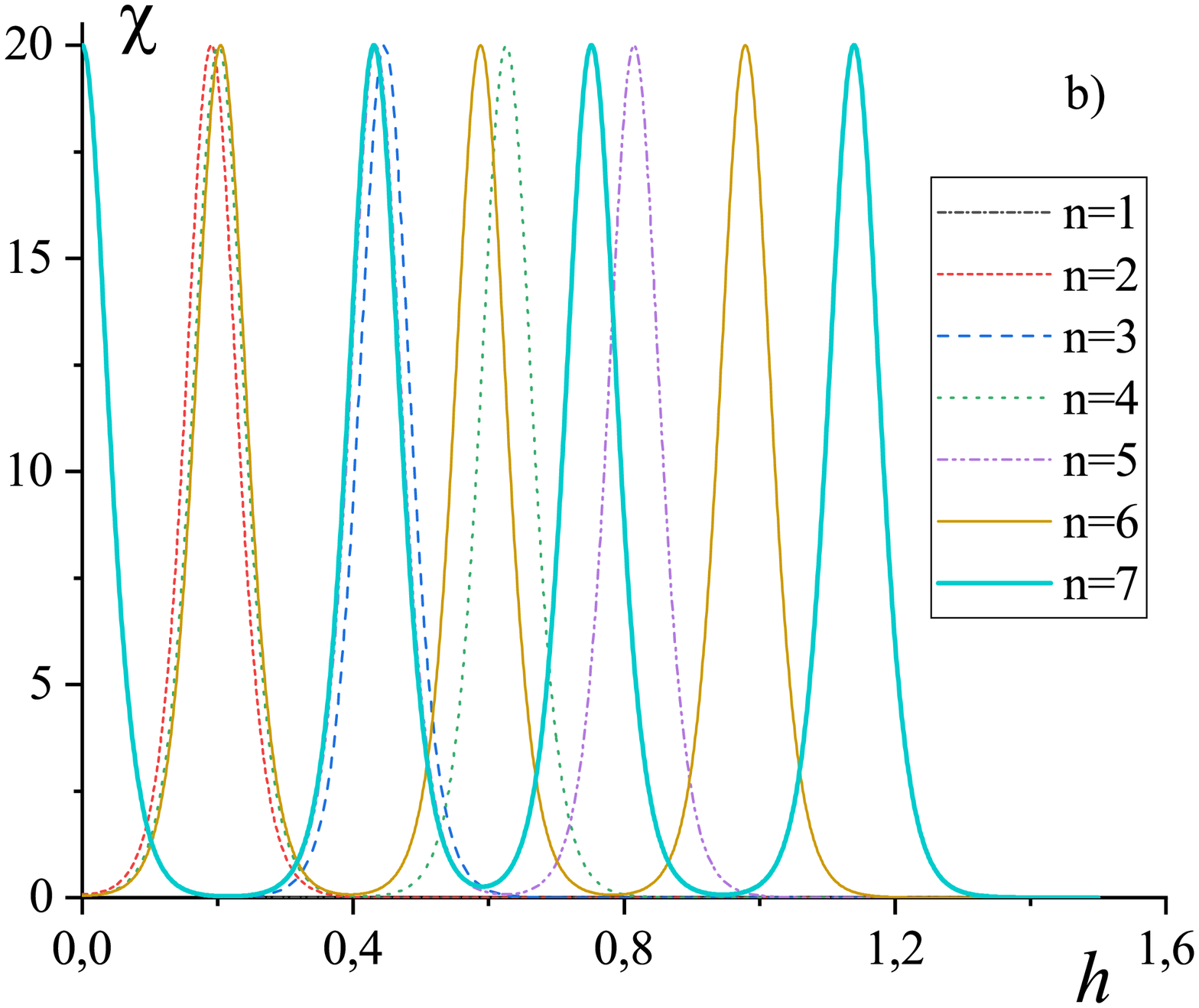}}
\centerline{\epsfysize=5.5cm\epsfbox{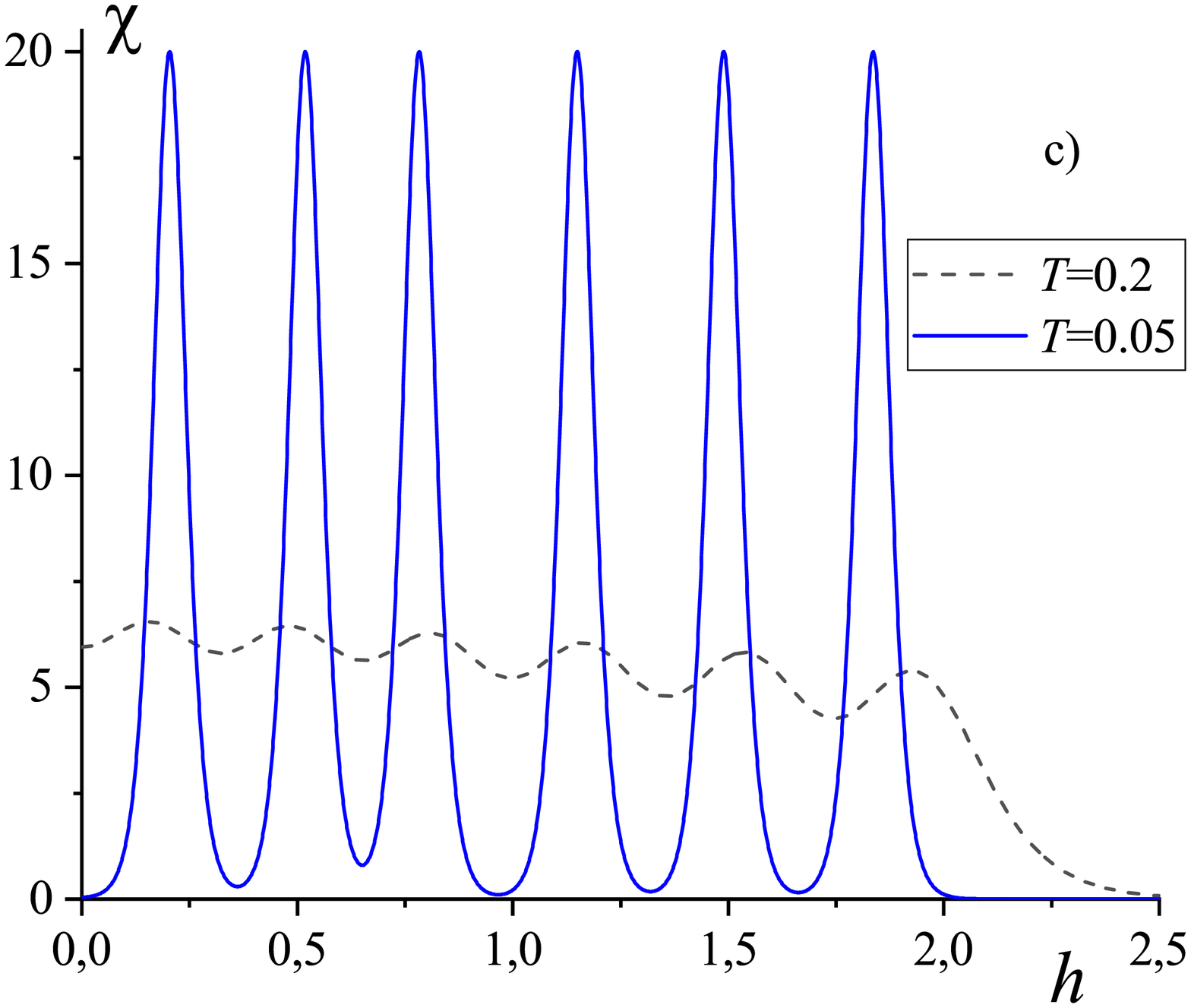}}
\caption{The magnetic spin susceptibility $\chi$ of linear clusters with $n$ electrons {\it versus} magnetic field at relatively high temperature $T=0.2$ (a) and low temperature $T=0.05$ (b). The comparison of $\chi(h)$ (c) for $T=0.2$ (black, dashed) and $T=0.05$ (blue, solid) for $n=12$.}\label{fig3}
\end{figure}

The magnetic spin susceptibility $\chi(h)$ of linear 2DWCs is demonstrated in Fig.~\ref{fig3}. The dependence contains peaks arising from the sequential dispairing of spins, starting from spins with a weak bonding at the ends of the chain. At large temperatures the peaks merge. At $h=0,~T\to 0$, $\chi(h)\to 0$ for an even $n$ and $\chi(h)\to const$ (maximum) for an odd $n$. The explanation is the pairing of all spins in the case of even $n$ and pairing of all spins but one, in the case of odd $n$ that gives the single-electron spin susceptibility. If the temperature grows, the zero $\chi(h)$ converts to a minimum.
Unlike the linear 2DWC, the circular ones possess the shell structure. The electron in different shells have strongly different bonds to other shells. In the cases of 2-5 electrons, all of them compose a single shell with the symmetry towards the $i\to j$ replacement. The difference with the linear cluster is the cyclic shape of the cluster. However, the oscillations occur due to the partial pairing.

Starting from $n=6$ two or more shells appear. The single electron in the center at $n=6$ or $n=7$ remains dispaired.

\begin{figure}
\centerline{\epsfysize=5.5cm\epsfbox{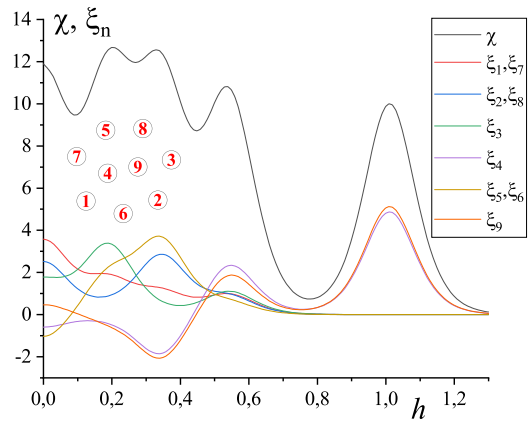}}
\caption{The magnetic spin susceptibility $\chi(h)$ of a round 2DWC with 9 electrons and mean spin of $k$-th electron derivatives $\xi_k(h)$ {\it versus} magnetic field. Inset. The structure of a round 2DWC with 9 electrons. The numbers inside the circles numerate the electrons. The peaks of $\chi(h)$ can be attributed to specific electrons.}\label{fig4}
\end{figure}

Round 2DWCs (see Fig.~\ref{fig4}) have a much more complicated spin ordering than linear ones. If the distance between shells exceeds the interelectron distance inside the shell, the spin ordering between inner and outer shells occurs at a much lower $h$ than the ordering in the shell. However, if the shell contains an odd number of electrons, the dispaired electron in one shell can be paired with the electron in another shell, but with a smaller pairing energy. So, the pairing of distant electrons can be superimposed in the simple picture of a single shell, producing a more complicated peak structure of $\chi(h)$. We see such behavior in Fig.~\ref{fig4}.

Note that more complicated variants of pairing are possible. In particular, the direct exchange of two distant spins is very weak. In that case the indirect pairing prevails \cite{anderson}. An example of such situation is presented in Fig.~\ref{fig4}, where the electrons experience the transition between the antiferromagnetic ordering at $h<J_{ij}$ to the ferromagnetic one at $h>J_{ij}$.

\section{Spin correlation function}
The spin structure of the Wigner cluster can be effectively elucidated by the spin correlation function $\zeta_{i,j}=\langle\sigma_i\sigma_j\rangle$, where $\langle ... \rangle$ denotes statistical averaging. The infinite Wigner lattice is homogeneous. Hence, $\zeta_{i,j }$ depends on the relative distance between spins. In the Ising lattice, at $h=0$, the phase transition occurs at some temperature $T_c$. Near $T_c$ the correlation radius infinitely grows and the correlation function becomes power-like. This behavior determines the scaling laws for the susceptibility at $\tau=(T_c-T)/T_c\to 0$ and $h\to 0$. In the infinite system the 2nd order phase transition is accompanied by the infinite length correlation. The 2D triangular lattice experiences a single phase transition. No phase transition occurs in a finite magnetic field. The susceptibility $\chi$ in a magnetic field gradually depends on $h$. On the contrary, in the 2DWC, the multiple reconstructions occur when $h$ changes, and that results in the periodical susceptibility $\chi$ dependence on $h$.

Unlike the infinite Wigner lattice, the cluster is not homogeneous. Hence, $\zeta_{i,j}$ depends on the coordinates of 2 spins. It is obvious that the scaling behavior will remain, to some extent, in a non-uniform system like a cluster, when the correlation length becomes comparable with the cluster radius. However, the absence of translational long-range lattice order and the cluster finiteness will destroy the phase transition. The strong exponential sensitivity of $J_{ij}$ to the interelectron distance essentially affects the correlation function.

The general shared feature of different clusters is the pairing of the most strongly connected nearest electrons if there is an even number of them. They compose ''molecules''. Electrons which have no pair in this close community are paired with more distant ones. If there are a couple of such electrons, they share an unpaired spin. However, in the system with many electrons different kinds of pairing are mixed.

\begin{figure}[ht]
\centerline{\epsfysize=2.3cm\epsfbox{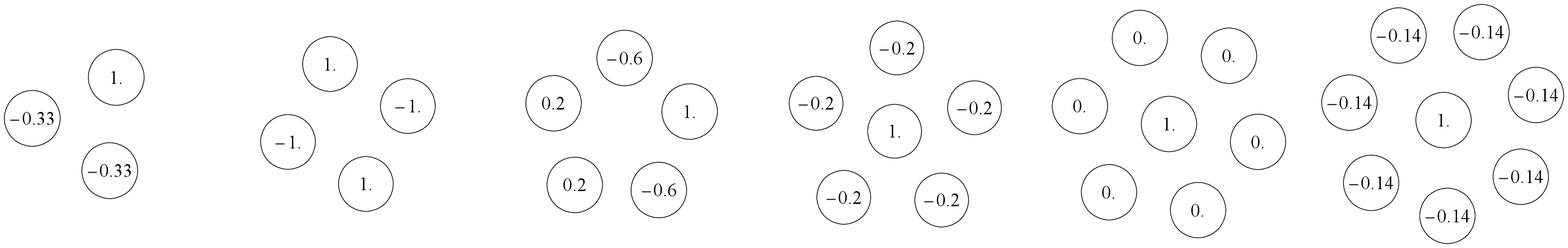}}
\vspace{0.25cm}
\centerline{\epsfysize=3.15cm\epsfbox{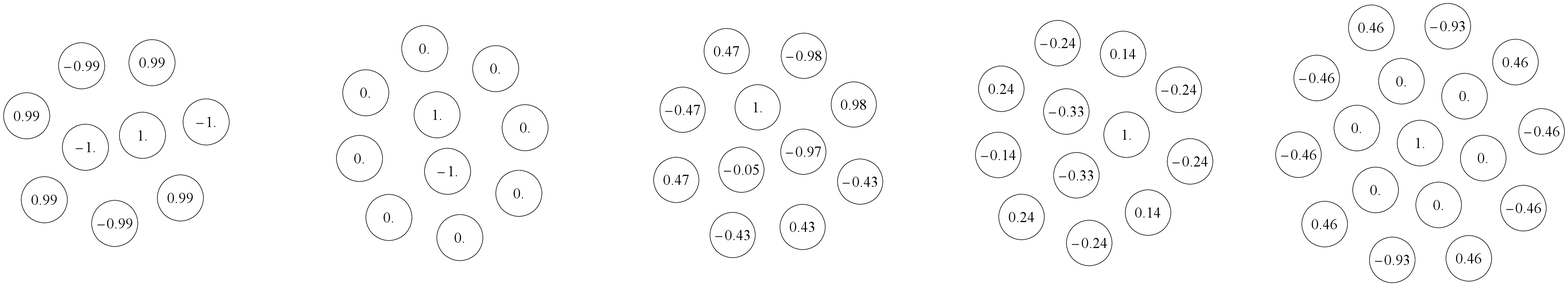}}
\caption{Correlation function $\zeta_{i,j}=\langle\sigma_i\sigma_j\rangle$ of spins for different clusters with 3-12 and 17 electrons. The spatial structure of clusters corresponds to the minima of Hamiltonian Eq.~(\ref{f0}). For number $i$ we chose the electron with the smallest $r_i$. The value of $\zeta_{i,j}$ is shown inside the circle denoting the corresponding electron located in the corresponding place. The limit $h\to 0$ $\beta\to\infty$ is presented. The antiferromagnetic ordering is seen.}\label{fig5}
\end{figure}

In Fig.~\ref{fig5} are the values of $\zeta_{i,j}$ for $i$ chosen to have the lowest value of $r_i$. We set the zero temperature limit to emphasize the antiferromagnetic ordering of spins. The correlation function has its infinite radius in the linear clusters. In the circular clusters the correlation drops with distance $|{\bf r}_i-{\bf r}_j|$. The drop follows from statistical arguments. The non-ideality of the crystallographic lattice also contributes to the correlation function decay.

\onecolumngrid

\twocolumngrid

Below we analyze the dependence of $\zeta_{i,j}$ for clusters with a small electron number. At zero temperature in the linear clusters the quantity $\zeta_{i,j}$ does not decay. The fixing of one spin forces nearby spins to have the opposite value. In circular clusters, the $\zeta_{i,j}$ behavior becomes more complicated. The weak interaction between shells and a smaller distance between inner electrons make their spins stronger bound. The other spins adjust to inner shells. The systems with small electron numbers $n=3,4,5$ form regular polygons. In these polygons the situation depends on the parity of $n$.

Let us analyze the obtained correlation functions. At zero $h$, systems with an even electron number tend to the pairing of all electrons, while, in the odd-electron system, one electron remains unpaired.

In the system with 2, 3, 4 and 5 electrons, the particles are located equidistantly. All positions are equivalent. One can readily count all configurations with the same lowest energy and find the correlation functions (see Fig.~\ref{fig5}).

The 2DWCs with 2 and 4 electrons at $h=0$ have their total polarization and all $\zeta_{1,i}=\pm 1$. Consider the case $n=3$ with the interaction between nearest neighbors. Let $i=1$ and fixed $\sigma_1=1$. The lowest energy is reached for configurations $(1,1,-1)$, $(1,-1,-1)$, $(1,-1,1)$. Averaging the product $\sigma_1\sigma_j$ we have $\zeta_{1,1}=1,~\zeta_{1,2}=\zeta_{1,3}=-1/3$. Analogically, for $n=5$ we obtain $\zeta_{1,1}=1,~\zeta_{1,2}=\zeta_{1,5}=-3/5,~\zeta_{1,3}=\zeta_{1,4}=1/5$.

For $n=6$, one electron is in the center, and the rest ones populate the vertices of a regular pentagon. The central spin is paired to a single spin on the outer shell. When distributed over 5 sites, one spin gives $\zeta_{1,i\neq 1}=-1/5$.

The cluster with $n=7$ contains 1 electron in the inner shell and 6 electrons, which are paired, in the outer shell. That gives $\zeta_{1,j}=0$.

The case of $n=8$ is similar to $n=6$: a single shared spin on the seven outer shell sites gives $\zeta_{1,j}=-1/7$.

If $n=9$, the system has lost the axial symmetry. Two central spins are strongly bound, $\zeta_{1,j}=-1$ for the inner electron and for the nearest to the external electron of inner electrons. Other external electrons have $\zeta_{1,j}\approx\pm 1$, conserving the alternation to minimize the total energy, except for two spins which could not fulfill the alternation.

A 10-electron cluster contains two electrons in the inner shell which spins are strongly bound, $\zeta_{1,j}=-1$, and 8 electrons, which are paired, in the outer shell. That gives $\zeta_{1,j}=0$.

The system with $n=11$ and $n=12$ also possesses central near-regular triangles. However, the spin ordering in these cases is strongly different. This can be explained by the system symmetry. The case $n=11$ has the axes of reflection, and the case $n=12$ has the $C_3$ symmetry.

For $n=11$, $\zeta_{1,j}$ is antisymmetric around the axes. As a result, one of quantities $\zeta_{1,j}$ in the inner shell is small, and the other one has values $\approx\pm 1$. One can also take into account the square on the right up of that plot. Similar to the case with $n=4$, for the best pairing, all spins in this square should be $\approx\pm 1$. The other spins have alternating values $\approx\pm 0.5$.

The inner shell of the 12-electron cluster has the spin ordering as the system with 3 electrons. This spin ordering forces the values of $\zeta_{1,j}$ for the outer shell.

Thus, the correlation function behavior becomes totally understandable qualitatively.

Now analyze the cluster with $n=17$. This cluster contains shells with 1, 6 and 10 electrons. The second and the third shells form an approximately regular hexagon and decagon. This figures have mutual elements of symmetry: two orthogonal reflection planes and the rotation on angle $\pi$, thus, obeying the group $C_{2v}$. The even number of electrons in the second shell results in their total pairing with $\zeta_{1,j}$ for $j$ within this group. The symmetry yields the separation of the electrons on the external shell onto 3 subsets with 2, 4 and 4 electrons, which have identical $\zeta_{1,j}$ values. This is just the case in the computed result. The subset with 2 elements is closer to the origin, then the other one is almost totally paired with the element $i$. The other third shell electrons are partly oriented by the elements of the subset with 2 elements and have alternating spins.

\section{Spin structure reconstructions {\it versus} spin susceptibility. Spin pairing spectroscopy}
The presence of multiple peaks allows one to attribute them to specific spin pairs. This attribution can be called ''spin-pairing spectroscopy''. The peak value of $h$ corresponds to the bonding strength. In a simple case this quantity coincides with $J_{ij}$ for specific pair $\{i,j\}$. However, if the distance between electrons $r_{i,j}$ is large, the exponential drop of $J_{ij}$ makes the direct exchange too weak. In this case, the indirect action along some chain of reorienting spins, which connect $i$ with $j$, can result in a stronger power interaction between $\sigma_i$ and $\sigma_j$. Here we analyze different clusters at different fields to systematize the spin ordering kinds.

For example, consider the round cluster with $n=9$ electrons, two of which are located inside the cluster at a small distance, and others compose the external shell with near-same distances between them. The corresponding dependence $\chi(h)$ is shown in Fig.~\ref{fig4}. To attribute the peaks, we plot also the $h$- derivative of mean individual spin. We see that all $\chi(h)$ peaks can be attributed to definite spins and spin pairs. The inner electrons are close together. Then the peak at $h=0$ is connected with an unpaired electron located on the outer shell. This electron is easier oriented by weak $h$. It results in the maxima of spin derivatives $\xi_n(h)=d\langle\sigma_n(h)\rangle/dh$ for shell electrons $n=1,2,7,8, 9$. At the strongest $h$ there is only one peak at $h\approx 1$, resulting from the strongest bound pair of the 4th and 9th electrons. The peak at $h=0.2$ can be attributed to the 3rd electron (and, partially, to the minima for the 2nd and 8th ones). The peak at $h=0.35$ is mostly connected with the peaks for the spins of the 5th, 6th, 2nd and 8th electrons and the minima for the 4th and 9th ones. The origination of these peaks can be understood from the geometric cluster structure. Thus, the magnetic field spectrum allows studying the spatial cluster organization.

Note that, as can be seen in Fig.~\ref{fig4}, the pairing-dispairing occurs not necessarily with the pair of closest electrons. It also reflects on the spins of other electrons (see maximum at $h=0.2$, for $\xi_3$, which is accompanied by minima for $\xi_2$ and $\xi_8$ and a shoulder for $\xi_5$ and $\xi_6$).

\subsection{Estimations of the characteristic parameters}
We expect that the most pronounced 2DWC manifestation can be obtained in the systems with a relatively small dielectric constant and a large electron mass. Besides, the material purity is important to exclude potential fluctuations localizing the holes. To approach the experimental situation we have estimated the parameters, characterizing the 2DWC for different systems, where we hope the electron spatial and spin  ordering are essential. The most suitable systems are electrons on the liquid He surface or free suspended p-type semiconductor layers. To exclude the Fermi energy the charge density is chosen low enough. However, this lower the limiting temperature of the 2DWC melting. We can not strongly lower the electron concentration, because the exchange interaction exponentially drops with the inter-electron distance. Consider electrons on the He surface, p-type GaAs channel, and 2D material MoS$_2$. It is convenient to chose the exchange integrals to be several times lower, than unit.

Let us estimate the characteristic parameters. Let $n=12$ (Figs.~\ref{fig2} and \ref{fig3}c), $k=1.66\cdot10^{-2}$~meV/cm$^{2}$, $\epsilon=1$ (free suspended layer). Then $R=6.18\cdot10^{-3}$~cm, $n_s=10^5$~cm$^{-2}$, the energy and temperature unit $E_0=0.26$~meV~$=2.99$~K, and the length unit $L=5.58\cdot10^{-4}$~cm. The dimensionless temperatures in the plots 0.05 and 0.2 correspond to $T=0.15$~K and $T=0.6$~K. Corresponding maximal exchange energies $J_{ij}$ are 0.043~meV (He), 0.11~meV (GaAs), 0.093~meV (MoS$_2$). The characteristic magnetic fields at $h=1$ are $B=2.23$~T (He), $B=1$~T (GaAs), $B=1.43$~T (MoS$_2$). These values look extremal enough, but achievable.

\section{Conclusions and Discussion}
We have studied the spin momentum, correlation function and the magnetic susceptibility of 2D Wigner clusters with linear and round shapes. It is found that the magnetic momentum dependence on the magnetic field differs from the Curie-Weiss law being almost linear up to the saturation value at relatively large temperatures. The low-field low-temperature magnetic susceptibility is zero for even and finite for an odd electron number, and that is explained by the spin pairing.

The magnetic susceptibility oscillates with the finite magnetic field due to a sequent disparing starting with the weakest and ending with the strongest spin bonds. At low temperatures the dependence consists of sharp peaks. In a linear chain, the average susceptibility per electron is of the order of $1/n$.

In the linear cluster the number of oscillations is equal to half the number of electrons due to pairing. Susceptibility is predominantly determined by unpaired electrons. This statement correlates with the average susceptibility value.

For the round cluster, the susceptibility oscillations are less regular. This is due to the shell structure of the cluster (weak bonding between the electrons of different shells), the approximate periodicity of the electron arrangement on the shell and the cyclicity of an individual shell (absence of ends). As a result of the shell structure, different shells act almost like independent subsystems. The periodicity of the electron arrangement in a shell makes all electrons equally subject to the magnetic field, but intershell pairing leads to the irregularity of magnetooscillations.

Compare a 2D Wigner cluster with the infinite lattice. The infinite Wigner lattice is homogeneous. Hence, the correlation function depends on the relative distance between spins. In the Ising lattice at $h=0$, the phase transition occurs at some temperature $T_c$. Near $T_c$ the correlation radius infinitely grows and the correlation function becomes power-like. This behavior determines the scaling laws for the susceptibility at $ \tau=(T_c-T)/T_c\to 0$ and $h\to 0$. It is obvious that this scaling behavior will remain, to some extent, in a non-uniform system like the 2DWC, when the correlation length becomes comparable with the cluster radius.

Unlike the infinite Wigner lattice the cluster is not homogeneous. Hence, the correlation function depends on the coordinates of two spins.

In the infinite system the 2nd order phase transition is accompanied with the infinite length correlation. The 2D triangular lattice experiences a single  phase transition. No phase transition occurs in a finite magnetic field. The susceptibility $\chi$ in a magnetic field gradually depends on $h$. On the contrary, in the 2DWC, multiple reconstructions occur when $h$ changes, and that results in the periodic susceptibility $\chi$ dependence on $h$.

Note that the magnetic field action can be replaced by the cluster rotation with angular frequency $\Omega=\frac{eB}{mc}$. The 2DWC in the axially symmetric potential is capable of free rotation without a change of shape or destruction \cite{weJETPL} that gives rise to a possibility to use the rotation instead of the magnetic field.

\paragraph*{\bf Acknowledgments.} This research was supported by the RFBR, grant No. 20-02-00622.

\section{Appendix. When is the Ising model more correct than the Heisenberg model?}
The Ising model is based on the replacement of the quantum spin operator $\hat{\bm s}_i$  of spin 1/2 by the  classical $c$-number $\sigma_i$. This reduces the $2^n$-dimensional Gilbert spin state to the space of $n$-dimensional binary numbers $\sigma_i$, reducing the NP-complete problem to NP-noncomplete. From the physical point of view, this can be justified when, instead of 2x2 Pauli matrices, the Hamiltonian will include only one component of them, say, $\sigma={\hat s}_z$. This is the case when the spin Hamiltonian has the form:
$$\sum_{ij}J_{ij,\alpha\beta}{\hat s}_{i\alpha}{\hat s}_{j\beta},$$
where $\alpha$ and $\beta$ stand for the Cartezian coordinates $(x,y,z)$, and $J_{ij,zz}\gg J_{ij,xx},~J_{ij,zz}\gg J_{ij,yy}$. If so, only diagonal components of spin, $\pm 1$ are included and the Heisenberg Hamiltonian converts to the Ising Hamiltonian.

However, the symmetry allows such a case in simple models $J_{ij,\alpha\beta}=J_{ij}\delta_{\alpha\beta}$. In that case one can not convert the Heisenberg model to the Ising one. Then, the Ising approach is only the approximation of a more general Heisenberg model.

Nevertheless, there are systems where this is not the case. Imagine the hole Hamiltonian in a usual cubic semiconductor like GaAs. In the bulk isotropic case, this Hamiltonian reads:
$$H=Ap^2+B({\bf pj})^2,$$
where $\bf j$ is the operator of spin 3/2. The surface quantization splits the states $m=\pm 3/2$ and $m=\pm 1/2$ so that 3/2 usually becomes the top heavy-hole state. Double degenerate $m=\pm 3/2$ states are analogical to the $\pm 1/2$ state of free electron with the exception that now the exchange integral $J_{ij,\alpha\beta}\approx J_{ij}\delta_{\alpha z}\delta_{\beta z}.$ In fact, the exchange interaction Hamiltonian reads:
$$H_{ex}\propto ({\bf j}_{i,\alpha}{\bf j}_{j,\beta}).$$
Projecting ${\bf j}_{i,\alpha}$, ${\bf j}_{j,\beta}$ onto the subspace of states with $j_z=\{3/2,-3/2\}$ we see that $\langle\pm3/2|j_x|\pm3/2\rangle=0$, $\langle\pm3/2|j_y|\pm3/2\rangle=0$, $\langle3/2|j_z|3/2\rangle=3/2$. This diagonalizes the Hamiltonian and  converts it to the Ising one. Oppositely, for the states with $j_z=\pm 1/2$, $\langle\pm1/2|j_x|\pm1/2\rangle \neq0$, $\langle\pm1/2|j_y|\pm1/2\rangle \neq0$, and this situation should be described by the Heisenberg model.

Then, such holes are good candidates to produce the Ising-Hamiltonian-controlled 2D hole Wigner clusters.

\end{document}